\documentclass[12pt]{article}
\usepackage[pdftex]{graphicx}
\usepackage{amssymb}
\usepackage{amsmath}

\newcommand{\rf}[1]{(\ref{#1})}
\newcommand{\beq}{\begin{equation}}
\newcommand{\eeq}{\end{equation}}
\newcommand{\bea}{\begin{eqnarray}}
\newcommand{\eea}{\end{eqnarray}}

\newcommand{\e}{\mbox{e}}
\renewcommand{\d}{\mbox{d}}

\newcommand{\lam}{\lambda}

\newcommand{\ra}{\rangle}
\newcommand{\la}{\langle}

\newcommand{\cD}{{\cal D}}

\newcommand{\cH}{{\cal H}}

\newcommand{\hH}{{\hat{H}}}

\newcommand{\omitt}[1]{ () }

\newcommand{\tinyspace}{\hspace{0.0556em}}
\newcommand{\dbltinyspace}{\hspace{0.1112em}}

\newcommand{\negdbltinyspace}{\hspace{-0.1112em}}

\newcommand{\bra}[1]{ \langle #1 | }
\newcommand{\ket}[1]{ | #1 \rangle }
\newcommand{\vac}{ \bra{{\rm vac}} }
\newcommand{\cuum}{ \ket{{\rm vac}} }

\begin{document}

\begin{center}
\vspace{24pt}
{ \large \bf Wormholes, a fluctuating cosmological constant and\\ the Coleman mechanism}

\vspace{30pt}

{\sl J. Ambj\o rn}$\,^{a,b}$,
{\sl Y. Sato }$\,^{c,d}$,
and {\sl Y. Watabiki}$\,^{e}$

\vspace{48pt}
{\footnotesize

$^a$~The Niels Bohr Institute, Copenhagen University\\
Blegdamsvej 17, DK-2100 Copenhagen \O , Denmark.\\
email: ambjorn@nbi.dk\\

\vspace{10pt}

$^b$~IMAPP, Radboud University,\\ 
Heyendaalseweg 135,
6525 AJ, Nijmegen, The Netherlands

\vspace{10pt}

$^c$~ Institute for Advanced Research, Nagoya University,\\
Chikusaku, Nagoya 464-8602, Japan\\  
{email: ysato@th.phys.nagoya-u.ac.jp}
\vspace{10pt}

$^d$~Department of Physics, Nagoya University\\ 
Chikusaku, Nagoya 464-8602, Japan\\

\vspace{10pt}

$^e$~Tokyo Institute of Technology,\\ 
Dept. of Physics, High Energy Theory Group,\\ 
2-12-1 Oh-okayama, Meguro-ku, Tokyo 152-8551, Japan\\
{email: watabiki@th.phys.titech.ac.jp}

\vspace{10pt}

}
\vspace{48pt}
\end{center}


\begin{center}
{\bf Abstract}
\end{center}

We show that in a two-dimensional model of quantum gravity the summation over all possible wormhole configurations 
leads to a kind of Coleman mechanism where the  cosmological constant plays no role for large universes. Observers who are 
unable to observe the change in topology will naturally interpret the measurements of the size of the universe as being  
caused by a fluctuating cosmological constant, rather than fluctuating topology of spacetime.

\noindent 
\vspace{12pt}
\noindent \\
\medskip
\vspace{12pt}
\noindent \\
PACS: 04.60.Ds, 04.60.Kz, 04.06.Nc, 04.62.+v.\\
Keywords: quantum gravity, low dimensional models, lattice models.

\newpage

\section{Introduction}\label{introduction}

In a quantum theory of gravity it has always been a problem how to think of the change of topology of spacetime. Should
such changes be allowed in the quantum theory of gravity? Presently, we do not know if a theory of four-dimensional
quantum gravity exists {\it an sich}, or whether it has to be a part of a large theory, like string theory. It is even difficult 
to discuss precisely what the change of topology is supposed to mean, since if we talk about differential manifolds
one cannot even in principle classify such topologies. Of course one could restrict by hand the class of manifolds and geometries
such that  a discussion makes some sense, or one could argue that manifolds and geometries are only derived, approximate
long distance  concepts coming from a different underlying theory, maybe again from string theory.  However, even in string theory 
we are facing this problem. String theory can be viewed as two-dimensional quantum gravity coupled to certain matter fields, and 
in this case we know, expanding around a fixed spacetime background, that we, by the requirement of unitarity, have to 
sum over all two-dimensional topologies of the string. Thus viewed as a two-dimensional theory of gravity, it is a gravity theory 
 where we {\it  have to} sum over topologies. We do not  know how to do this in an unambiguous way. Such a summation is 
usually divergent and no physical principle has so far told us uniquely how to perform such summation. This is even true 
in the toy models called non-critical string theories, although in these theories one  can formally in many cases perform a summation 
over genera and find a non-perturbative partition function. In this article we will consider an even simpler ``string field theory''
where it is also possible to perform the summation over all genera, and we will show that from the point of 
view of ``one-dimensional observers'' such a world, where spatial universes can split and merge, appears as a world 
with a fluctuating cosmological constant. Also, the model provides an explicit example where one can test the so-called
Coleman mechanism, which loosely states that summing over all wormholes of a theory of quantum gravity the effective cosmological
constant must be zero. As we will see this is true, but maybe not the way Coleman had in mind.  
The rest of the article is organized as follows: in  Sec. \ref{two} we describe the
so-called CDT string field theory and the result of summing over all topologies. In Sec. \ref{three} we show that some of the results of the CDT string field theory can be obtained in a simple way by allowing the cosmological ``constant'' to fluctuate. Finally Sec.\ \ref{four} contains a discussion of the results.
 
\section{CDT string field theory}\label{two}

Causal dynamical triangulations (CDT) is an attempt to provide a non-perturbative lattice regularization  of quantum gravity. 
In the model one can rotate to spacetimes of Euclidean signature and the four-dimensional model can be studied by computer 
Monte Carlo methods. Some interesting results have been obtained (see \cite{physrep,loll} for reviews). Here we will be 
interested in the two-dimensional toy model. It can be solved analytically, and the continuum limit, where the lattice spacing 
goes to zero can be taken \cite{al,ab}. The continuum theory, which can be shown to correspond to  a particular sector
 \cite{agsw} of two-dimensional Horava-Lifshitz gravity \cite{horava}, is simple, 
 and it describes the propagation of a closed one-dimensional spatial universe of length $\ell(t)$
as a function of the proper time $t$ used. The Hamiltonian is given by   
\beq\label{3.8}
\hH_0(\ell) = -\ell \frac{\d^2}{\d \ell^2} +\lam \;\ell.
\eeq
This is a standard Hermitian operator on wave functions $\psi(\ell)$
on the positive real axis, which are 
square-integrable with respect to the scalar product
\beq\label{3.9}
\la \psi_1 | \psi_2\ra = \int_0^\infty \frac{\d \ell}{\ell}\;\psi^*_1(\ell) \psi_2(\ell).
\eeq                                                                                                                                                                                                                                                                                                                                                                                                                                                                                                                                                                                        
The amplitude of a universe which starts out at time $t=0$ having spatial volume (length) $\ell$, i.e.\ in a quantum state $| \ell\ra$,  
will evolve (in Euclidean time) according to
\beq\label{ja1}
G_0(\ell,\ell'; t) = \la \ell' | \e^{-t \hH_0(\ell)} | \ell \ra.
\eeq  
A complete set of eigenfunctions for $\hH_0$ with corresponding eigenvalues are of the form 
\beq\label{ja2}
\psi_n^{(0)}(\ell) = p_n(\ell) \; \e^{- \sqrt{\lam} \; \ell},\quad E_0(n) = 2n \sqrt{\lam}, \quad n=1,2,\ldots,
\eeq 
where $p_n(\ell)$ is a polynomial of order $n$ and $p_n(0) = 0$.  In particular, $p_1(\ell) = 2\sqrt{\lam}\, \ell$. The so-called Hartle-Hawking wave function is defined as
\beq\label{ja3}
w_0(\ell) = \int_0^\infty \!\!\d t\;G_0 (\ell, \ell'=0; t) = \e^{-\sqrt{\lam} \; \ell},\qquad \hH_0(\ell)  w_0(\ell) =0.
\eeq
Formally the Hartle-Hawking wave function $w_0(\ell)$ is an eigenfunction of $\hH_0$ with eigenvalue $E_0(0) = 0$. However,
it is not normalizable according to the definition \rf{3.9}.  The Hartle-Hawking wave function $w_0(\ell)$ 
describes a universe which starts out in the state $| \ell\ra$ and eventually disappears. The evolution of a
universe, governed by \rf{ja1} is such that if the universe starts out as a circle of length $\ell$, the topology of space 
will not change. The spatial universe cannot split in two and it cannot merge with another universe. 
\begin{figure}[h]
\vspace{-1cm}
\centerline{\scalebox{0.25}{\includegraphics{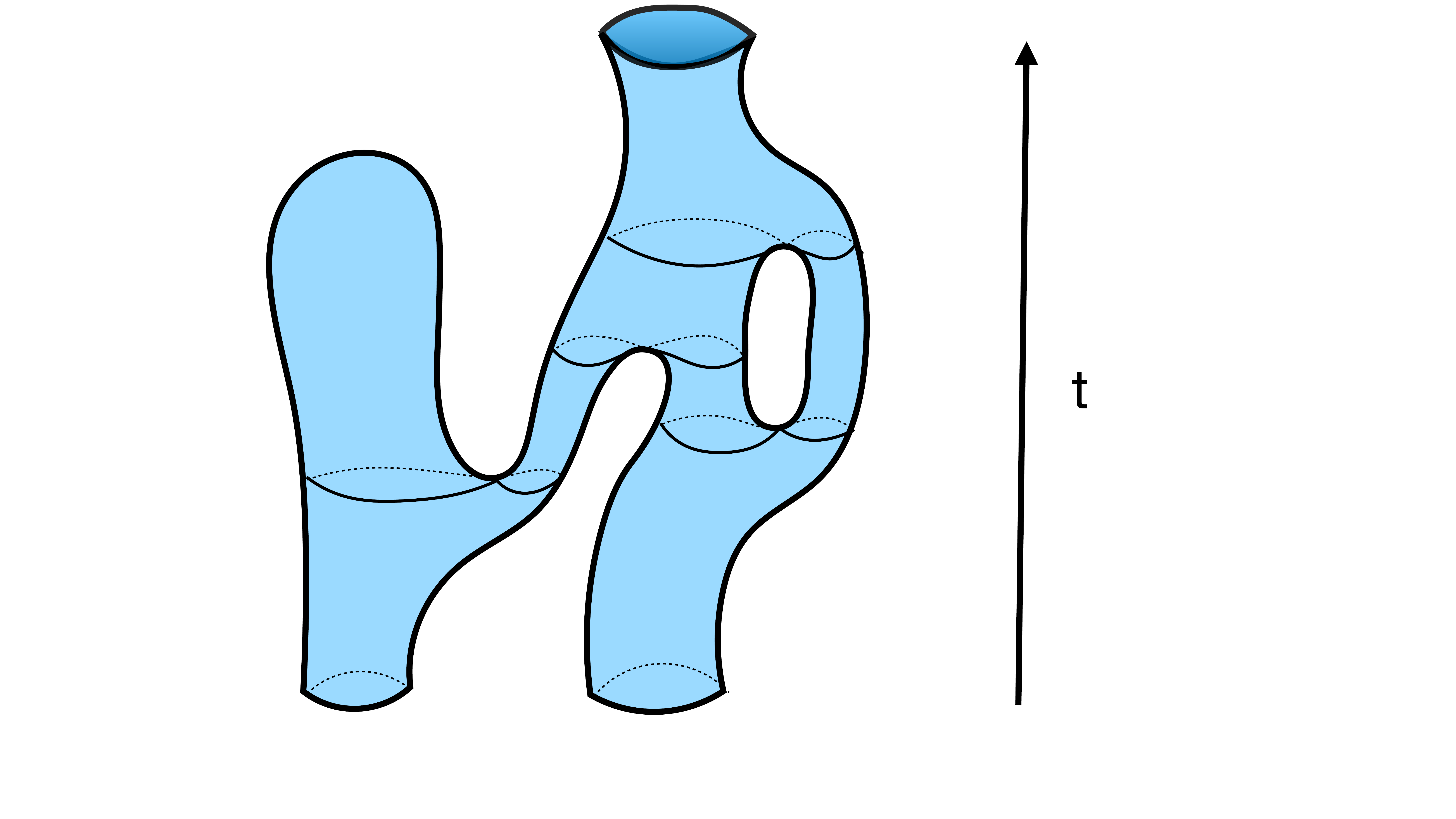}}}
\vspace{-1cm}
\caption{{\small Dynamics of two spatial universes according to \rf{Hgeneral}. 
At time $t=0$ we have two spatial universes, universe 1 and 2. Both propagate 
according to the first term in \rf{Hgeneral}. At a later time the left universe (universe 1) splits into two according to the third term in 
\rf{Hgeneral}. The left of these again propagates according to the first term, but then disappears in the vacuum, 
a process made possible by the the second term (a tadpole term) in \rf{Hgeneral}. Universe 2  likewise splits in two, of which the left part merges with the right split of universe 1, a process made possible by the fourth term in \rf{Hgeneral}. 
Then one of these universes merges with the universe at the right of the picture, according to the fourth term in \rf{Hgeneral}.
Finally this universe merges with the right split of universe 2. During this time evolution the number of spatial universes at a given 
time changes from 2 at $t=0$, to 3 at the first split, to 4 at the second split and then to 3 and 2 at the following mergers and finally 
to one when the left universe disappears in the vacuum.}}
\label{fig0}
\end{figure}
The possibility to generalize this picture, such that baby universes were created was considered \cite{gcdt}, and it was shown to be connected to a certain matrix model (it even defined a new continuum limit of matrix models) in \cite{matrix}.  In \cite{cdtsft} 
it was developed into a full CDT string field theory, or a third quantization in the terminology of quantum gravity, where spatial 
universes can split and merge, governed by a ``string coupling constant'' $g$. 
More precisely, one introduces a ``multi-universe'' Fock space, where 
$\Psi^{\dagger}(\ell)$ creates one closed spatial universe of length $\ell$ with a marked point, 
and $\Psi(\ell)$ annihilates one closed spatial universe with no marked points (the marking and not-marking of these spatial loops 
are just for combinatorial convenience when solving the models).
The commutation relations of the universe (or string)  operators are 
\bea
&&
\big[\, \Psi(\ell) \,, \Psi^\dagger(\ell') \,\big]
\,=\,
\delta(\ell-\ell')
\, ,
\label{CommutePsi}
\\
&&
\big[\, \Psi(\ell) \,, \Psi(\ell') \,\big]
\,=\,
\big[\, \Psi^\dagger(\ell) \,, \Psi^\dagger(\ell') \,\big]
\,=\,
0
\, .
\label{ZeroCommutePsi}
\eea
The vacuum states (no string states), in analogy with the no particle states in
many body theory, are denoted $\cuum$ and $\vac$ and  
are the Fock states defined by
\beq
\Psi(\ell) \cuum \,=\, 
\vac \Psi^\dagger (\ell) \,=\, 
0
\, .
\eeq
The string field Hamiltonian acting on this Fock space is 
\bea\label{Hgeneral}
\cH \!&=&\!
\int\limits_0^\infty\! \frac{d \ell}{\ell}  {\dbltinyspace}
  \Psi^\dag(\ell) {\tinyspace} \hH_0(\ell)
  \,
  \ell\Psi(\ell) - \int\limits_0^\infty\! d \ell {\dbltinyspace}
  \delta(\ell) {\tinyspace}\Psi(\ell)
\nonumber\\&&\!
-\; g
\!\int\limits_0^\infty\! d \ell_1
\!\int\limits_0^\infty\! d \ell_2 {\dbltinyspace}
  \Psi^\dag(\ell_1) {\tinyspace}
  \Psi^\dag(\ell_2) {\tinyspace}
  (\ell_1{\negdbltinyspace}+{\negdbltinyspace}\ell_2)
  \Psi(\ell_1{\negdbltinyspace}+{\negdbltinyspace}\ell_2)
\nonumber\\&&\!
-\;  {\tinyspace} g
\!\int\limits_0^\infty\! d \ell_1
\!\int\limits_0^\infty\! d \ell_2 {\dbltinyspace}
  \Psi^\dag(\ell_1{\negdbltinyspace}+{\negdbltinyspace}\ell_2) {\tinyspace}
  \ell_1 \Psi(\ell_1) {\tinyspace}
  \ell_2 \Psi(\ell_2)
\,.
\eea
In this theory one can in principle  calculate
any amplitude for a universe to merge, split or disappear in the vacuum. The power of $g$ will indicate the total number 
of splitting and merging and one can develop a perturbation theory, where $G_0(\ell,\ell';t)$ acts as  the free propagator.
We have illustrated a process in Fig.\ \ref{fig0}.
In particular it is possible to consider the amplitude where $n$ spatial universes of length $\ell_1, \ell_2, \ldots, \ell_n$ present 
at $t=0$ evolve in all possible ways and eventually disappears. The connected part of this amplitude is a generalization 
of $w_0(\ell)$, which we denote $w(\ell_1,\ldots,\ell_n)$. In \cite{cdtsft} it was shown that this amplitude is precisely the 
so-called $n$-loop amplitude of the  matrix model defined in \cite{matrix} and in \cite{allorder} 
it was shown that  $w(\ell_1,\ldots,\ell_n)$ can be found non-perturbatively and can be expressed
entirely in terms of the one-loop amplitude $w(\ell)$. Further, in \cite{stochastic-cdt} it was shown 
that the results to all orders in the string coupling 
constant $g$ could be described as by a change of the Hamiltonian $\hH_0$ to a new ``non-perturbative'' Hamiltonian\footnote{In 
eqs.\ \rf{Hgeneral} and \rf{ja4} the coupling constant $g$ is assumed to be positive. The choice $g \geq 0$ goes all the way back
to the very formulation of GCDT \cite{gcdt}, where $g$  was the coupling constant related to the creation of baby universes in CDT,
and where it was always assumed that $\ell \geq 0$. In a more abstract setting of string field theory it might be convenient to drop 
such restrictions (see \cite{aw1} for a discussion), and i.e.\ eq.\ \rf{ja5} is invariant under $g,\ell \to -g,-\ell$. However, here we will always assume $g, \ell \geq 0$.}
\beq\label{ja4}
\hH_0(\ell) = -\ell \frac{\d^2}{\d \ell^2} +\lam \;\ell \quad \to \quad \hH(\ell) = -\ell \frac{\d^2}{\d \ell^2} +\lam \;\ell -g \ell^2.
\eeq
In particular, we have for the Hartle-Hawking wave function 
\beq\label{ja5}
  \hH_0(\ell)  w_0(\ell) =0  \quad  \to  \quad \hH(\ell)  w(\ell) =0,
  \eeq
  where ( {\it Ai} and {\it Bi} denote the standard Airy functions)
  \beq\label{ja6}
   \quad w_0(\ell) = \e^{-\sqrt{\lam} \; \ell} \quad \to \quad w(\ell) = 
   \frac{{\rm Bi}\left(\frac{\lam -g \, \ell}{g^{2/3}}\right)}{{\rm Bi}\left(\frac{\lam}{g^{2/3}}\right)} 
   +c \cdot {\rm Ai} \left(\frac{\lam-g \, \ell}{g^{2/3}}\right).
\eeq
 It is seen that $w(\ell)$ has an asymptotic expansion in $\xi = g/\lam^{3/2}$, starting with $w_0(\ell)$. This 
 is illustreted in Fig.\ \ref{fig1}. 
 \begin{figure}[t]
 \vspace{-1cm}
\centerline{\scalebox{1.0}{\includegraphics{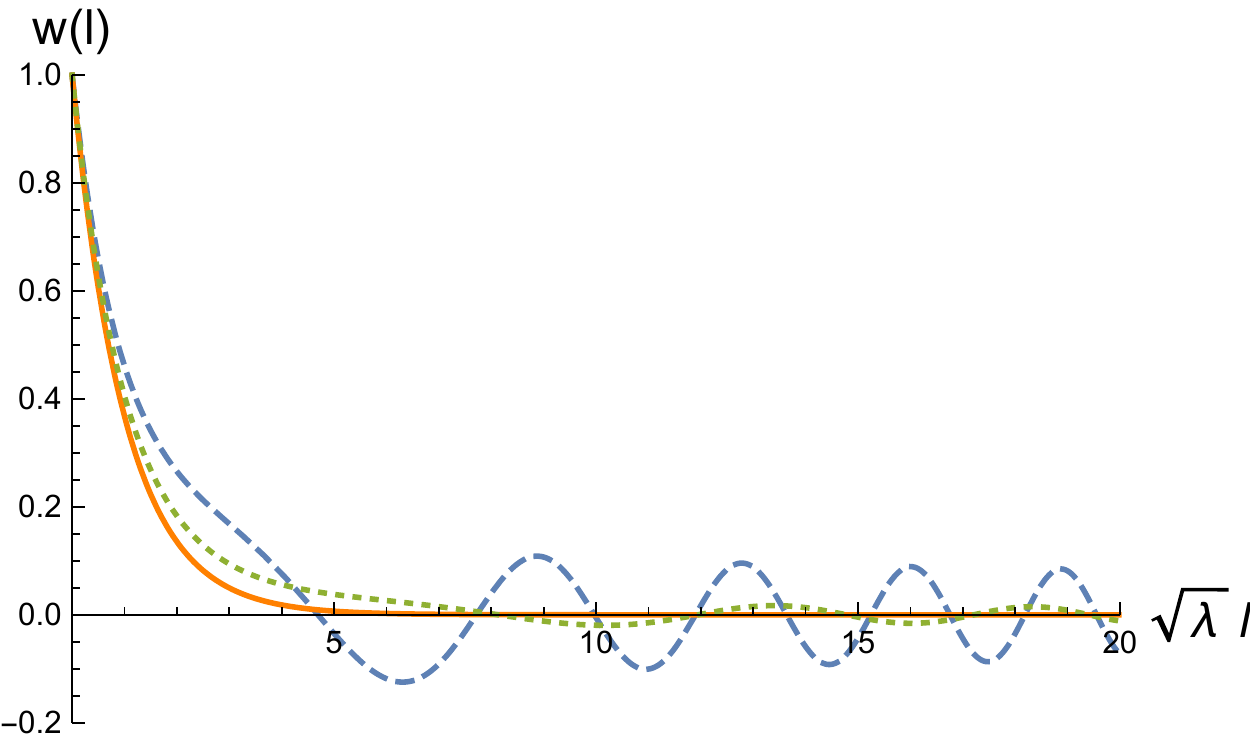}}}
\caption{{\small $y$-axis: $w_0(\ell)$ (orange, continuous line) and $w(\ell)$ with $c=0$ for $\xi =1/6$  (green dotted line) 
and $\xi = 1/3$ (blue dashed line). $x$-axis: $\sqrt{\lam}\, \ell$.}}
\label{fig1}
\end{figure}
 The first terms agree with the ones calculated perturbatively using CDT string field theory.
 $c$ is an arbitrary constant which is not determined by the perturbative expansion in $g$ (which is all we know 
 from first principle). 
 
 The Hamiltonian $\hH$ is unbounded from below and it has a one-parameter of self-adjoint extensions which still have 
 a discrete spectrum $E(n)$ with corresponding eigenfunctions $\psi_n(\ell)$ (see \cite{ak} for a discussion of this 
 in a similar situation in Liouville gravity and the ``ordinary'' double scaling limit of matrix models). 
 A WKB analysis shows that as long as $E_n < \sqrt{\lam}/(4\xi)$, $E(n)  \approx E_0(n)$ and in addition $\psi_n(\ell)$ is well 
 approximated by $\psi^{(0)}_n (\ell)$ provided $  \sqrt{\lam}\, \ell <   1/\xi$. For $\sqrt{\lam}\,  \ell > 1/\xi$ the exponential fall off of $\psi_n^{(0)}(\ell)$
 is replaced by an oscillatory behavior where $\psi_n (\ell)$ only falls off like $1/\ell^{1/4}$ (in agreement 
 with the behavior of $w(\ell)$ from \rf{ja6}). This implies that while 
 the spatial extension of a typical universe at a given time $t$ for a state $\psi_n^{(0)}(\ell)$ was of order $1/\sqrt{\lam}$, 
 the spatial extension of a typical universe for the state $\psi_n(\ell)$ will be infinite. In addition this dramatic change is 
 caused by infinite order $g$ contributions. To any finite perturbative order in $g$ we would still find a spatial extension of 
 order $1/\sqrt{\lam}$.  From a CDT string field point of view this change of behavior thus  comes from infinite order change of 
 the spatial topology. If we naively assume that ``we'', as one-dimensional observers, can measure   $\la \ell(t) \ra$, 
 we should see quite a difference depending on wether the evolution of the universe is governed by $\hH_0$ or by $\hH$.
 
 \section{A fluctuating cosmological constant}\label{three}
 
 The quantum mechanics corresponding to the hamiltonian $\hH_0$ is reproduced by the path integral 
 \beq\label{ja7}
 G_0(\ell,\ell';t) = \int_{\{\ell(s)\}} \cD \ell(s) \; \e^{-S[ \ell (s) ]}, \qquad \cD \ell(s) = \prod_{s=0}^{s=t} \frac{\d\ell (s)}{\ell(s)}, 
 \eeq
 where the integral is over functions $\ell(s)$ where $\ell(0) = \ell$ and $\ell(t) = \ell'$, and where 
 \beq\label{ja8}
 S_0[\ell(s)] = \int_0^t \d s \; \Big( \frac{\dot{\ell}^2}{4 \ell(s)} + \lam \ell(s) \Big),\qquad  \dot{\ell} = \frac{\d \ell(s)}{\d s},
 \eeq
 (see \cite{ai} and \cite{agsw} for details).
 Assume now that the cosmological coupling constant is not really constant, but a variable which performs Gaussian fluctuations
  in time $t$ around the value $\lam$ with a standard deviation $\sigma = 2\sqrt{g}$. This would change the propagator
  in \rf{ja7} to 
  \bea
  G(\ell,\ell';t) &=&   \int \cD \nu(s) \cD \ell(s)\; 
  \exp \left[ - \int_0^t \d s \Big(\frac{\dot{\ell}^2}{4 \ell(s)} + (\lam + \nu(s)) \ell(s)   + \frac{1}{4g} \nu^2(s) \Big)\right]\nonumber \\
  &=& \int \cD \ell(s)\; \exp \left[ - \int_0^t \d s \Big(\frac{\dot{\ell}^2}{4 \ell(s)} + \lam \ell(s)   - g \;\ell^2(s) \Big)\right],
  \label{ja9}
 \eea
 where we recall that we have assume $g \geq 0$.
 The resulting  unboundedness of the  effective potential $\lam \ell - g \ell^2$ can be traced to the fact that even for a small 
 standard deviation $\sigma = 2\sqrt{g}$  there is a non-zero probability that the cosmological term $\lam + \nu(t)$ 
 can be negative, resulting in a term which is unbounded from below when $\ell \to \infty$. 
 It is also seen from the effective Lagrangian appearing in \rf{ja9} that for  the functional integral to be  well defined, the
 boundary conditions on $\ell (t)$ at infinity has to such that the kinetic term  counteracts the unboundedness of the potential.
 Such a behavior at $\ell \to \infty$ is precisely the one we described in Sec.\ \ref{two}.
We can now determine the quantum Hamiltonian corresponding to the propagator \rf{ja9} by standard methods as in \cite{ai}, and we find
\beq\label{ja10}
G(\ell,\ell';t) = \la \ell' | \e^{-t \hH(\ell)} | \ell\ra,\qquad \hH(\ell) =  -\ell \frac{\d^2}{\d \ell^2} +\lam \;\ell -g \ell^2,
\eeq
i.e.\ precisely the Hamiltonian \rf{ja4} derived from CDT string field theory. We thus have the remarkable situation that 
a ``trivially'' fluctuating gravitational constant produces a result identical to the summation over all possible merging and splitting 
of space in a third quantized theory of quantum gravity. This result was observed before, where it was interpreted in the 
contest of stochastic quantization \cite{kawai,stochastic-cdt}, 
but here we want to change the perspective as discussed in the next section.

\section{Do wormholes matter for your universe?} \label{four}

According to what is known as Coleman's mechanism \cite{coleman}, 
we owe our whole existence to wormholes and baby universes (for a review, see, e.g. \cite{Hebecker:2018ofv})\footnote{Coleman's mechanism was discussed in the Lorentzian context in \cite{Kawai:2011rj, Kawai:2011qb}.}. 
The possibility that 
our universe can create so-called baby universes only weakly connected to our universe or  create wormholes connecting
different parts of the universe is responsible for the smallness of the cosmological constant according to the Coleman 
mechanism. This ``prediction'' of the consequences 
of quantum fluctuations of geometry has not really been tested since we so far lack a model of quantum gravity which allows
for sufficient detailed calculations in four dimensions. However, the theory of two-dimensional quantum gravity has a different 
status. It is a renormalizable theory, it allows for the creation of baby universes and wormholes, both in the case of 2d Euclidean 
quantum gravity (Liouville quantum gravity)  and the case of Horava-Lifshitz gravity. Both quantum theories 
can be derived as  scaling limits of lattice regularized versions of the theories, and the fact that these regularized versions 
of the theories can be solved combinatorially allows us to perform the summation over all topologies  and study the effect of 
such a summation (see \cite{book} for a review of the combinatorial view on these models). 
As mentioned above CDT is  the lattice regularization of two-dimensional Horava-Lifshitz gravity,  
and it allowed to solve some aspects of the complete third quantization of the gravity theory, defined by the string field 
Hamiltonian $\cH$ given in eq.\ \rf{Hgeneral}. Let us now discuss what the solution tells us about Coleman's mechanism.
We can use the Hartle-Hawking wave function $w(\ell)$  as a starting point. It is not normalizable, but this is a short distance problem
and we will be interested in large $\ell$.
Alternatively we could use one of the energy eigenstates 
for $\hH$, which is normalizable. As mentioned the expansion parameter is $\xi =g/\lam^{3/2}$ 
and as long as $\sqrt{\lam} \, \ell \ll 1/\xi$ the relative probability distribution of $\ell$ dictated by $w(\ell)$ follows that of 
$w_0(\ell) = e^{-\sqrt{\lam} \,\ell}$. However for $\sqrt{\lam}\, \ell \geq 1/\xi$ it is drastically different, to the extent that $w(\ell) $ does 
not fall off exponentially but only as $1/\ell^{1/4}$. In fact the large $\ell$ behavior depends on the dimensionless variable 
$\sqrt{g} \,\ell^{3/2}$, rather than on $\sqrt{\lam} \, \ell$. Thus we can say that the large scale structure of the universe is 
independent of the cosmological constant $\lam$ and it not governed by the exponential behavior associated with $\lam$. 
In this sense Coleman's mechanism {\it is} working: the cosmological constant is irrelevant for the large scale structure. What
is not quite like the simplest version of Coleman's mechanism is that the resulting large scale structure is not simply given 
by a classical background where one puts the cosmological constant to zero. The infinite genus universes seem to play 
a crucial role, and this is true no matter how small the expansion parameter $\xi$ is. Eventually, for $\sqrt{\lam}\,\ell \gg 1/\xi$ infinite 
genus surfaces dominate the behavior of $w(\ell)$ and all the eigenfunctions $\psi_n(\ell)$ of $\hH$.  

Let us image that we are one-dimensional beings living in a universe where $g=0$. The Hamiltonian is thus $\hH_0$ 
and if the universe is in the lowest energy eigenstate state $\psi_1^{(0)} (\ell)$ 
then a number of measurements of the spatial volume (length)  
$\ell(t)$ of our universe at times $t_1,t_2,\ldots$ will result in a distribution 
$P^{(0)}_1(\ell) =   \big( \psi_1^{(0)} (\ell)\big)^2/\ell= 4\lam\, \ell \,\e^{-2\sqrt{\lam} \ell} $. If one had a model for the universe, stating
that the quantum evolution is governed by $\hH_0$ and that the universe is in the energy state corresponding 
to the eigenvalue $E_0 (1)$ of $\hH_0$, then the 
observations would allow us to determine the cosmological constant $\lam$. If, in reality, $g >0$, but $\xi \ll 1$ and 
the universe was in the energy state $\psi_1(\ell)$ of $\hH$, then as long 
as we can only measure $\ell$s where $\sqrt{\lam}\, \ell\ll 1/\xi$ we would reach the same conclusion. 
However, as our measurements improved 
and we could measure larger and larger values of $\ell$, we would observe disagreement with the predictions coming from 
$\hH_0$. Assuming that we were never able to actually observe a change in topology, we would be tempted to 
conclude that our observations support the idea that the cosmological constant is actually not constant, but 
fluctuating like $\lam(t) = \lam + \nu(t)$. That assumption would result in an effective, but weird Hamiltonian $\hH$, unbounded 
from below and it would explain the observed distribution  $P_1(\ell) =  \big( \psi_1 (\ell)\big)^2/\ell$ of measured $\ell$.
Still one would most likely say that a picture provided by the string field Hamiltonian $\cH$, given by \rf{Hgeneral} is 
much more satisfactory, and it allows potentially for a detailed description of the dynamics of merging and splitting of universes.
It also calls for a better understanding of what kind of infinite genus spacetime will actually appear and be important in our model.
This question can actually be addressed because the model is so simple, and preliminary results 
\cite{toappear} indicate that these spacetimes
are in some sense ``nice'' infinite genus spacetimes \cite{infinitegenus}. We believe that research in this direction 
could be important both for ``real'' string theory and for models where two-dimensional CDT is used to construct higher dimensional universes \cite{aw1,aw}. As an example,  let us mention that it follows from the models considered in \cite{aw1,aw} 
that if the wavelength
of a typical oscillation of the cosmological constant is  larger than a couple of billion years, our theory suggests that the present 
acceleration of our universe can be explained by the summation over baby universes and wormholes in CDT, in this way unifying 
physics at the shortest distances and at the largest scales.  Details of such an estimate will appear in a separate publication.

\section*{Acknowledgement}
The work of YS was supported by Building of Consortia for the Development of Human Resources in Science and Technology 
and by JSPS KAKENHI Grant Number 19K14705. 
The work of YW was partially supported by JSPS KAKENHI Grant no. JP18K03612.

\end{document}